# Oscillation and collective conveyance of water-in-oil droplets by microfluidic bolus flow

Takuya Ohmura,[1] Masatoshi Ichikawa,[1] Ken-ichiro Kamei,[2] and Yusuke T. Maeda[1,3,4,*,#]

[1]*Department of Physics and Astronomy, Graduate School of Science, Kyoto University, Oiwake-cho, Kitashirakawa, Kyoto 606-8502, Japan*

[2]*Institute for Integrated Cell-Material Sciences (WPI-iCeMS), Kyoto University, Yoshida-Ushinomiya-cho, Kyoto 606-8501, Japan*

[3]*The Hakubi Center for Advanced Research, Kyoto University, Yoshida-Ushinomiya-cho, Kyoto 606-8501, Japan*

[4]*PRESTO, Japan Science and Technology Agency, 4-1-8 Honcho, Kawaguchi, Saitama 332-0012, Japan*

Microfluidic techniques have been extensively developed to realize micro-total analysis systems in a small chip. For microanalysis, electro-magnetic forces have generally been utilized for the trapping of objects, but hydrodynamics has been little explored despite its relevance to pattern formation. Here, we report that water-in-oil (W/O) droplets can be transported in the grid of an array of other large W/O droplets. As each droplet approaches an interspace of the large droplet array, while exhibiting persistent back-and-forth motion, it is conveyed at a velocity equal to the droplet array. We confirm the appearance of closed streamlines in a numerical simulation, suggesting that a vortex-like stream is involved in trapping the droplet. Furthermore, more than one droplet is also conveyed as an ordered cluster with dynamic reposition.

[*]Correspondence: ymaeda@phys.kyushu-u.ac.jp

[#]Present address: Department of Physics, Faculty of Sciences, Kyushu University

Recent developments in the fabrication of miniaturized devices have permitted the creation of micro-total analysis systems (μTAS) in the fields of bioengineering, systems biology, and single cell genomics [1,2]. These small devices can perform many types of biological and chemical analyses, from trapping and sorting of cells to droplet PCR [3,4], using only tiny amounts of solution. In addition, their low production costs allow them to be used disposably for diagnostic purposes in clinical medicine [5]. Typical devices are composed of a circuit of microchannels, solenoid valves, microchambers, and other functional modules such as electrodes, a configuration which enables highly accurate and high-throughput microanalyses [6,7]. Some of the most indispensable modules are those that realize the efficient separation and trapping of particles. Although optical tweezers can trap particles, its local weak force makes it difficult to handle in a fast flow. To overcome this difficulty, the S-shaped microchannel can separate differently sized particles owing to the interplay between shear-induced stress and inertial lift force and eventually align the sorted particles in a line [8,9]. In addition, an interfacial tension gradient across two laminar flows has been used [10,11]. The use of hydrodynamic effects for particle manipulations has received much attention, but there is a less explored effect which is also relevant—namely, collective dynamics.

A group of micro-objects individually interacting is the simplest realization of a many-body system. In colloidal particles [12,13], charged plasma [14,15], red blood cells in capillaries [16,17], micro-droplets [18-20] or motile living cells [21,22], ordered motion or patterns are emerged owing to their interactions through while single element behaves randomly. These organized dynamics emerge through long-range interactions in passive many-body systems [23]. One relevant system for trapping in microfluidics is the one-dimensional (1D) array of water-in-oil (W/O) droplets, which consists of a lattice of W/O droplets in a line and is maintained under immiscible oil flow in a microchannel. In this droplet array, individual droplets interact hydrodynamically with each other, and it is known that, under a strong confinement by floor and ceiling plates, longitudinal wave propagation occurs [24-28]. On the other hand, the array of droplets is more remindful of a transport system, with the small objects being trapped in their grid and conveyed in a single direction. To realize a droplet conveyor using an array of other droplets, it is first necessary to understand hydrodynamics in a crystal composed of differently sized droplets dragged by immiscible oil flow. However, this subject has yet to be addressed.

In this Letter, we study the trapping and sequential transport of W/O droplets in a developed microfluidic device. We placed each droplet in one of the grids in a one-dimensional array of other larger W/O droplets, and found that the droplet was conveyed at the same velocity as the large droplets in the lattice. Interestingly, the droplet embedded in the lattice was periodically moved in a back-and-forth manner relative to the large droplets, while the large droplets were dragged at a constant velocity by the oil fluid. This periodic motion was not symmetric to and from adjacent large droplets. That is, the small droplet moved faster (slower) from (toward) the large droplet behind it. To account for the trapping of droplets and oscillation, we performed a numerical simulation based on the lattice Boltzmann method. In the reference frame of large droplets, the parabolic profile of fluid velocity in the z axis and the reduced velocity of large droplets cause the streamlines to close. Furthermore, as more than one droplet was trapped in the grid of the large droplet array, these droplets assembled into a stable cluster with dynamic reposition and were conveyed at a velocity identical to that of the droplet array.

The microfluidic setup developed in this study is depicted in Fig. 1(a). Polydimethylsiloxane (PDMS) was cast on a silicone wafer in which the microstructure of SU-8 positive photoresist was patterned through conventional photolithography and then cured at 80℃ for 1 hour. The PDMS layer was cut carefully and then bonded with pre-cleaned slide glass by treating with oxygen plasma for 30 s. To strengthen the bonding of PDMS to the glass surface, we kept the PDMS devices at 80℃ for 1 hour. The inlet and outlet ports were punched-out with a biopsy punch and then the capillary tubes were connected. We used mineral oil supplemented with 2.0% Span80 surfactant and pure water so that the water-in-oil (W/O) droplets were generated by sheath flow. The experiment was carried using a velocity range of 1500~2700 μm/s for mineral oil and 500~1100 μm/s for W/O droplets. The flow rate of fluids was precisely controlled with a pressure-regulated pump system (MFCS System; Fluigent). Please note that the viscosity of fluids was 22 mPas for oil and 0.89 mPas for water, and the typical Reynolds number was $4.3~8.5 \times 10^{-3}$, indicating that inertia was negligible.

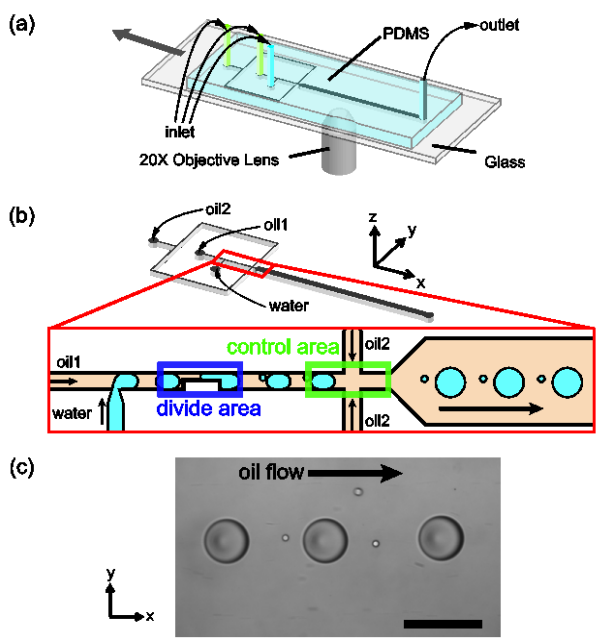

FIG. 1. Microfluidic setup
(a) A schematic of the microfluidic setup used in this study. A patterned PDMS was bonded onto slide glass. The PDMS microfluidic device was placed on a motorized stage in an inverted microscope. In order to observe the W/O droplets in a moving frame, we moved the motorized stage at a speed equal to the droplet motion but in the opposite direction. (b) Design of the microfluidic channel used in this study. While W/O droplet was generated at T-shaped junction, it was divided into two different sized droplets at divide area. In order to control the distance between two large droplets, we added another flow channel shown as control area. This module allowed us to control the distance of droplets without affecting the division of the W/O droplet. (c) A representative image of W/O droplets arranged in a line in an alternating sequence of large and small. The scale bar represents 200 μm.

We first investigated the dynamics of droplets arranged in a line in an alternating sequence of small and large. We generated successive W/O droplets at a T-shaped junction and then split one droplet into two differently sized droplets (Fig. 1(b)) [29,30]. In the divide area, the channel becomes narrow from 100 μm to 20 μm and the length of the narrow part was 350 μm. The small droplet was 10 to 40 μm in diameter and the large one was 100 to 200 μm in diameter. The one-dimensional array of different sized droplets were dragged by the surrounding oil flow in a microchannel whose width and entire length were 700 μm and 20 mm, respectively (Fig. 1(c)). The velocity of the large droplet was always slower than that of the oil flow owing to the friction with the ceiling and floor surfaces. One would expect that the small droplets would be moved by the oil at a comparable velocity because they are free from the frictional influence. However, we found that the small droplets exhibited a periodic back-and-forth motion relative to the large droplets (Fig. 2(a) and (b)), while the large droplets were conveyed at a constant velocity (Fig. 2(b), 1.2 mm/s for large droplets, oil flow of 2.7 mm/s). Neither longitudinal nor transverse waves were observed in this droplet array. In addition, after the small droplets turned away from the large droplets ahead of them, they continued to move astern until they came clos the large droplets behind them. This result implies that the small droplets were repelled by the large droplets, and eventually their repeated approaches and rejections resulted in an oscillatory motion. In addition, the observed oscillation could not be easily explained based on hydrodynamic interactions via a potential flow in the xy plane alone: According to the superposition of the dipole-like potential, the motion of droplets can propagate as a density wave in a droplet crystal but cannot have a stable orbit reminiscent of oscillation [24-28]. The approximation of a two-dimensional potential flow was no longer valid in this experiment, and therefore we needed to take the z-axis into account.

Next, we examined whether the small droplets were moved in the z-direction in the course of the back-and-forth motion. We can infer the z-position of small droplets by measuring their brightness, because drifting in the z-axis means defocusing, which alters the intensity of droplets. We found that the brightness of small droplets was periodically changed in the anti-phase of the oscillation of the droplet motion (Fig. 2(c)). The center of mass of small droplets was approximately 30 μm above the floor surfaces when the droplets moved backward relative to the large droplets. Moreover, we also analyzed the speed of small droplets in their forward and backward motions. As clearly seen in Fig. 2(d), the speed in the forward direction was greater than that in the backward motion. This asymmetric velocity was mostly observed in our experiments and it did not depend on the distance between large droplets, which suggests that hydrodynamic interaction with large droplets had little influence on the origin of the asymmetry. One possible mechanism to explain the speed asymmetry is the deceleration of the fluid flow when the small droplets moved close to the PDMS wall in their backward motion: The fluid of the oil phase sticks to the solid floor surface under a no-slip condition. Thus our results highlight the importance of considering the interaction among W/O droplets, oil fluid, and solid walls.

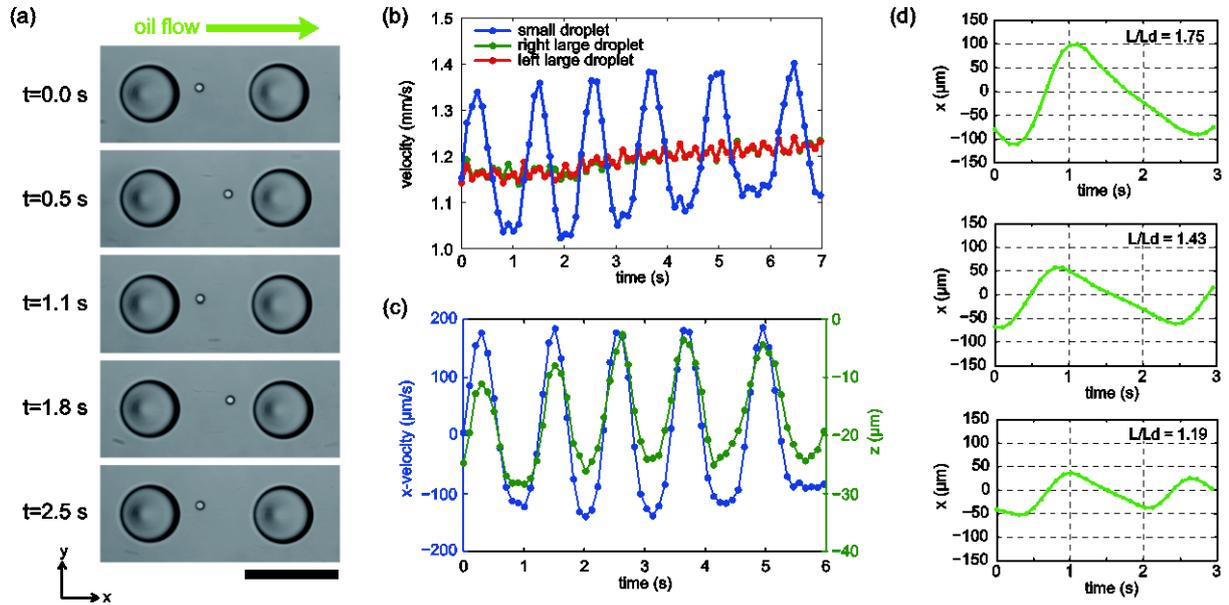

FIG. 2. Asymmetric, back-and-forth motion, and conveyance of droplets
(a) Snapshots of single small droplet between two large droplets (forward and backward). Small droplet moves in back-and-forth manner relative to the large droplets. (b) The dynamics of relative velocity. While large droplets move at a constant speed of 1.2 mm/s, the oscillation of relative velocity is clearly found for the small droplet. (c) The brightness of a small droplet, which reflects its z-position, shows anti-phase oscillation against its relative velocity. (d) The back-and-forth motion is asymmetric. The motion of a small droplet, relative to a forward-moving large droplet, in the forward direction is faster than that in the backward direction. $L$ represents the distance between large droplets and $L_d$ is the diameter of large droplets. The scale bar represents 200 μm.

These experimental results motivated us to propose a schematic model for the oscillation of single small droplets (Fig. 3(a)). In the reference frame of large droplets, a small droplet moves in the forward direction if it lies midway between the top and bottom walls. To test whether this model recapitulates the asymmetric oscillation of small droplets, we performed a numerical simulation based on the lattice Boltzmann method [31,32]. We performed the numerical simulation as follows: Two hard spheres were confined under two solid walls using a periodic boundary condition. The diameter of the large droplets was set at 120 µm, whereas the height of the microchannel was 123 µm. A pressure difference was applied in order to generate fluid

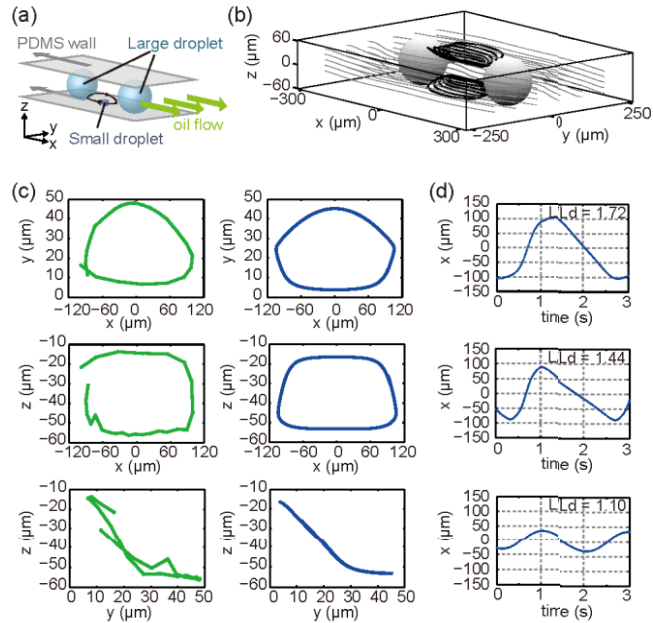

FIG. 3. Numerical simulation of microfluidic bolus flow

(a) A schematic of the simulated setup, showing the PDMS walls, large spheres and tracer particle. We assume that two large spheres whose diameter (120 µm) is comparable to the height of a channel (123 µm) are placed in the fluid flow. In addition, we also make the boundary of the channel, i.e., ceiling and bottom surfaces, move at a certain velocity in a direction opposite to the oil flow. (b) The streamline obtained in numerical simulation is shown in 3 dimensions. (c) Comparison of the trajectories of a small droplet observed experimentally (top) and those of a small sphere obtained by simulation (bottom). (d) Asymmetric back-and-forth motion of single small sphere.

flow at velocity $v = U_1$ in the laboratory frame. In order to determine the reference frame of large droplets, we had the top and bottom walls move in an opposite (backward) direction at $v = -U_2$, thereby the relative velocity of the oil fluid is $v = U_1 - U_2$, One small sphere of 20 μm was placed between two large droplets under a given set of initial conditions, and then its trajectory was analyzed. We note that no-slip boundaries existed among the oil phase, droplets and solid wall during the simulation because we added a sufficient amount of surfactant in the experiments.

Figure 3(b) shows the simulated streamlines of a fluid passing through large droplets. We found that the closed streamlines appeared between two large droplets. The results of a detailed comparison between the simulation and experiment are presented in Fig. 3(c). The simulated and experimental trajectories of a small droplet in a single period were strikingly similar, supporting the notion that such a droplet moves along a closed streamline. In addition, the small droplet moved faster in the forward direction in both the simulation and the experiment (Fig. 3(d)). The asymmetric nature of oscillation was always observed at $U_1 < U_2$, irrespective of the distance between large droplets. The ceiling and floor plates slowed down the surrounding fluid, as well as single small particle, at $U_1 - U_2$, and therefore asymmetric oscillation arose from the closed streamlines. We call this pattern of circulating streamlines a microfluidic bolus flow.

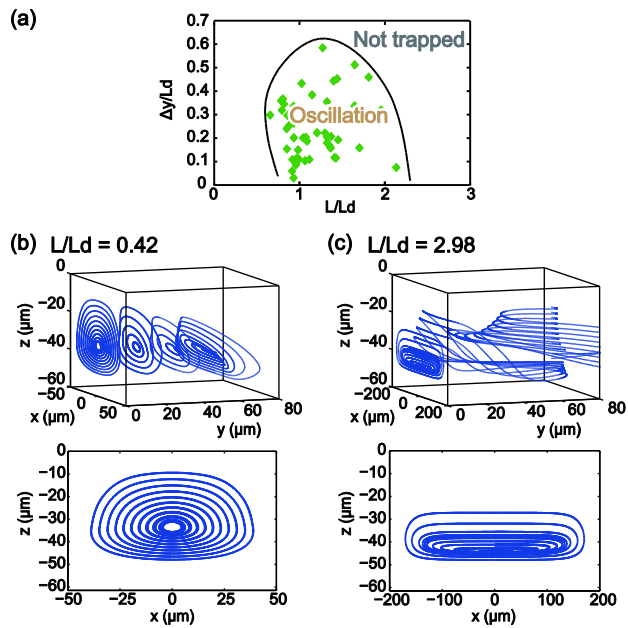

FIG. 4. Existence domain of microfluidic bolus flow

(a) A diagram of the trapping of small droplets with periodic motion. The horizontal axis is the distance between forward- and backward-moving large droplets, and the perpendicular axis is the deviation in the y-direction. (b and c) The streamlines in short ((b), $L/L_d$=0.42) and long ((c), $L/L_d$=2.98) interspaces of the droplet array. Data are shown in 3D (top) and for the zx plane (bottom).

Next, in order to project the region where the small droplets are conveyed while showing the periodic motion, we draw a diagram with two axes, one representing the distance between two large droplets, L, and the other the distance in the y direction, $\Delta y$ (Fig. 4(a)). In the experiment, we found that the optimal region was trapped with periodic motion at midst $L$ ($L_d/2 < L < 2L_d$) and small $\Delta y$ ($\Delta y < L_d/2$), whereas closed streamlines representing microfluidic bolus flow were always present irrespective of $L$ (the numerical simulation data are not shown here). To resolve the difference between experiment and simulation, we numerically examined the area of microfluidic bolus flow by changing $L$. As $L/L_d$ was increased in simulation, this bolus flow was distorted and the center of the vortex moved close to either the ceiling or floor plates (Fig. 4 (b) and (c)). Given that small droplets were placed in the middle of the z-axis for large $L/L_d$, the probability of being trapped in the microfluidic bolus flow could be decreased. In addition, $\Delta y$ also limited the observable region of the oscillation according to the relation $\Delta y < L_d/2$, which agreed with the numerical simulation (data not shown). In summary, microfluidic bolus flow existed over a wide range of our droplet array, but the initial position in the z-axis appeared to affect the efficiency of trapping.

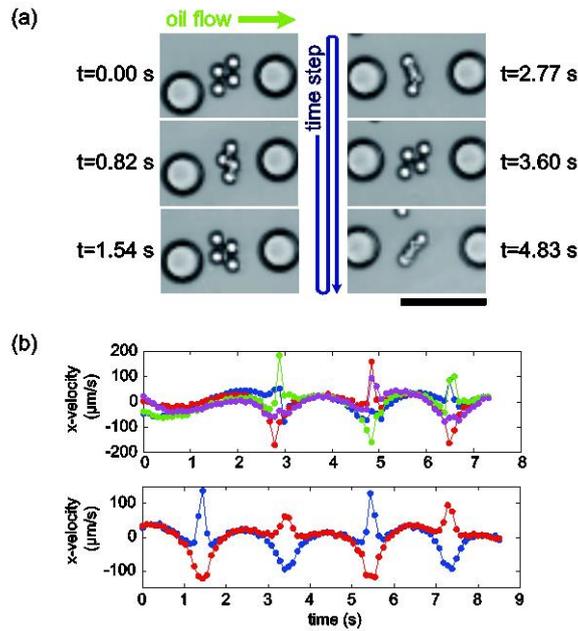

FIG. 5. Dynamic clustering of droplets

(a) The formation of a quadruplet cluster from four small droplets. Small droplets were readily trapped to form single cluster showing dynamic reposition. The scale bar represents 200 μm. (b) Plot of the relative velocity of small droplets in the quadruplet cluster (top) and doublet cluster (bottom) as a function of time. The time course of each droplet is shown in different colors. Increase and decrease of velocity reflects dynamic repositions inside a cluster.

Finally, we investigated a group of small droplets placed between two large droplets. We found that a quadruplet cluster was generated after four droplets were trapped simultaneously (Fig. 5(a)). The cluster was stably moved at the same velocity with large droplets, while exhibiting dynamic repositioning: Small droplets switched their position periodically while the four droplets maintained the cluster. The period of flipping was $3.6 \pm 0.2$ s. In addition, the formation of a doublet cluster was observed for two small droplets as well as the formation of a quadruplet cluster for four small droplets, indicating that more than one small droplet was trapped to form a dynamic cluster spontaneously in the grid of large droplet array. In addition, we plotted the velocity of droplets in a cluster with respect to time (Fig. 5(b)). The change of velocity at repositioning described a sharp rather than a smooth oscillation (Fig. 2(b)). The asymmetric shape of droplet clusters may affect the dynamics of flipping whereas groups of small droplets still maintain rotation in time, as do the single small droplets. An additional effect involved in the rotation of clustered droplets may be the parabolic profile of the laminar oil flow in the z direction. Once the stable cluster that can be assumed as a rigid body is formed, the fluid shear in the vicinity of the wall may cause rotational motion as seen in asymmetric Janus particles or yeast cells with a shmoo [33]. Detailed analyses of the effect of droplet shape remain to be conducted, as the hydrodynamic attraction needed to form an asymmetric cluster is beyond the focus of the present study.

In conclusion, we have shown the oscillatory motion of W/O droplets in a one-dimensional array of large droplets dragged by oil flow in a microfluidic channel. Single small droplets placed in the interspace of the droplet array were transported at the same velocity of the array while showing asymmetric oscillation by microfluidic bolus flow. The device in this study had a thickness of 123 μm and thus additional freedom in the z-direction led to dynamically organized motion such as the oscillation of single small droplets and even the clustering of multiple droplets. The bolus flow is thought to be a general phenomenon for which one can find many relevant biological and biomedical examples, such as closed streamlines between red blood cells in vessels [16,34,35]. This bolus flow may trap microvesicles involved in intercellular communications and then make clustered aggregations at the interspace of red blood cells for transport [36]. Further studies in other relevant systems, e.g., electrophoresis of the colloidal suspensions will contribute to a better understanding of the transport of particles in microfluidics [37].

**ACKNOWLEDGMENTS**

This work was supported by an internal grant from The Hakubi Center, an interdisciplinary collaborative grant from WPI-iCeMS, Kyoto University, and partial funding from Opto-Science Foundation and Japan Science and Technology Agency (PRESTO). We thank R. Murakami for assistance with the LBM simulation.